\documentclass[aps,prd,twocolumn,showpacs,amsmath,amssymb]{revtex4}
\usepackage{graphicx}
\usepackage{amsfonts,bm}
\usepackage[english]{babel}

\def\be{\begin{equation}}
\def\te{\end{equation}}
\def\ee{\end{equation}}
\def\ba{\begin{eqnarray}}
\def\bea{\begin{eqnarray}}

\def\tea{\end{eqnarray}}
\def\ea{\end{eqnarray}}
\def\eea{\end{eqnarray}}
\def\lla{\left\langle}
\def\rla{\right\rangle}

\begin{document}

\title{Macroscopic approximation to relativistic kinetic theory from a nonlinear closure}

\author{J. Peralta-Ramos}
\email{jperalta@df.uba.ar}
\author{E. Calzetta}
\email{calzetta@df.uba.ar}
\affiliation{Departamento de F\'isica, Facultad de Ciencias Exactas y Naturales, Universidad de Buenos Aires and IFIBA, CONICET, Cuidad Universitaria, Buenos Aires 1428, Argentina}


\begin{abstract}
We use a macroscopic description of a system of relativistic particles based on adding a nonequilibrium tensor to the usual hydrodynamic variables. The nonequilibrium tensor is linked to relativistic kinetic theory through a nonlinear closure suggested by the Entropy Production Principle; the evolution equation is obtained by the method of moments, and together with energy-momentum conservation closes the system. Transport coefficients are chosen to reproduce second order fluid dynamics if gradients are small. We compare the resulting formalism to exact solutions of Boltzmann's equation in 0+1 dimensions and show that it tracks kinetic theory better than second order fluid dynamics.
\end{abstract}
\maketitle

\section{Introduction}

Fluid dynamics successfully describes the long-wavelength, low-frequency dynamics of various systems including the quark-gluon plasma (QGP) created in ultrarelativistic heavy ion collisions \cite{revhydro1,revhydro2,revhydro3,stateart1,stateart2,stateart3,stateart4,libro}. It is usually linked to the more fundamental kinetic theory description by giving a closure, i.e., an expression for the one particle distribution function $f$ in terms of the hydrodynamic variables. The most widely used closures are those provided by the Chapman-Enskog expansion and the method of moments due to Grad (see e.g. \cite{cerc}), later on adapted to relativistic systems by Israel and Stewart \cite{is1,is2}. These two approaches rely on writing $f$ as a function of the same variables that describe the system at equilibrium, namely energy density $\epsilon$ and velocity $u^\mu$ for a system of relativistic massless particles as considered here. 

Traditional fluid dynamics derived from kinetic theory by these methods has two important limitations. First, it relies on an expansion in gradients of hydrodynamic variables, which necessarily implies that the system is sufficiently close to equilibrium so that these gradients are small. In turn, this means that the system is also very close to being isotropic in momentum space. Second, it breaks down at large shear viscosity-to-entropy ratio $\eta/s$. Comparison to relativistic kinetic transport calculations \cite{huov,eleta} shows that the applicability of Israel-Stewart (IS) theory to describe heavy ion collisions becomes marginal at $\eta/s > 0.2-0.3$, especially at very early times. 

The application of fluid dynamics to relativistic heavy ion collisions shows that the assumption of small gradients breaks down at the earliest times after impact, and that large momentum-space anisotropies actually survive for many fm/c. Besides, near the edges of the fireball created in these collisions the momentum-space anisotropies are large at all times \cite{martedges}.  These conclusions are also supported by studies based on the Anti de Sitter/Conformal Field Theory (AdS/CFT) correspondence, which show that viscous hydrodynamical behavior is reached at times when
the system still possesses large momentum-space anisotropies, and that these anisotropies persist during the entire evolution of the plasma \cite{adsEQ,adsEQ2}.

With respect to the second limitation of fluid dynamics mentioned above, one should bear in mind that what is actually extracted from data by comparison with viscous fluid dynamic simulations that use a temperature-independent $\eta/s$ is an average value of this ratio. In reality, $\eta/s$ depends  on the local temperature of the medium, dropping from 
high temperatures to a minimum at the critical temperature, and rising with decreasing temperature in the
hadronic phase \cite{eta,eta2}. In this phase, $\eta/s$ can reach values $\simeq 0.5-1.0$, which are too large for standard fluid dynamics to be reliable. Beyond this value of $\eta/s$, an
appropriate treatment of the dynamics of this matter should
switch smoothly from a viscous hydrodynamic description
to a kinetic one \cite{stateart1,stateart2,stateart3,stateart4,after1,after2}.

The purpose of this paper is to present a nonequilibrium effective theory (NET) capable of handling highly nonequilibrium situations 
and large values of $\eta/s$. The model was developed by the authors in \cite{dev} and applied to heavy ion collisions in \cite{app}. The link with kinetic 
theory was established in \cite{linking}, and the inclusion of color degrees of freedom was carried out in \cite{color} (we note that in this paper we do not consider colored particles). 

Here, we want to focus on the comparison of the NET to  second order fluid dynamics of \cite{hyd1,hyd2,hyd3} (we will call this theory SOFD, and note that it includes IS theory as a special case) and to kinetic theory. From numerical simulations of 0+1 boost invariant flow, we show that the NET tracks kinetic theory better than SOFD, even in the presence of very large momentum-space anisotropies or large values of $\eta/s$. Due to these features, the model can be useful to study in a unified way the dynamics of matter created in heavy ion collisions, from the early stage where large momentum-space anisotropies occur, through the deconfinement transition, beyond which large values of $\eta/s$ arise if the temperature dependence of $\eta$ and $s$ is taken into account, until freeze out.

In the context of heavy ion collisions, there have been previous extensions of standard fluid dynamics aiming at alleviating the above mentioned limitations (in other contexts there have been numerous extensions of standard fluid dynamics, see Section \ref{net}). In \cite{third}, an extension of IS formalism to third order in hydrodynamic gradients was carried out, and an approximate resummation to all orders was presented as well. It was found that the resummed theory tracked kinetic theory better that the third order formalism and much better than IS, even for (very) large values of $\eta/s \sim 3$.  In \cite{denicolnew}, the IS equations were rederived in a novel way based on the exact definition of the dissipative part of the energy-momentum tensor instead of on the truncated expression used by IS \cite{is1,is2}. This resulted in the IS evolution equations but with different expressions for the transport coefficients that greatly improved the agreement of this formalism with kinetic theory, even for large values of $\eta/s \sim 3$ as well. In \cite{mauricio,aniso,aniso2}, the method of moments was applied to an anisotropic distribution function instead of an isotropic one as it is usually done when deriving hydrodynamics from kinetic theory. The result was  a new formalism termed {\it anisotropic hydrodynamics} (AH) which can handle highly nonequilibrium situations and large values of $\eta/s$. We note that our NET is closer in spirit to AH than to the developments of \cite{third} and \cite{denicolnew}, because both the NET and AH include a nonhydrodynamic variable in addition to the usual hydrodynamic ones. As it is discussed later, this nonhydrodynamic variable accounts for the backreaction of the distribution function on the evolution of the hydrodynamic variables.

This paper is organized as follows. In Section \ref{net} we describe the entropy principle on which the NET is based, review the closure obtained from this principle and present the evolution equations. In Section \ref{comp} we compare the results of numerical simulations of the NET, SOFD and kinetic theory. This section contains our main results. In Section \ref{summ} we end up with a brief summary. 

\section{Nonequilibrium effective theory}
\label{net}

In this section we briefly review the NET that was developed in \cite{dev,linking,color} (see also \cite{app}).

The NET is obtained from the closure picked up by the so-called Entropy Production Principle (EPP) (a recent review can be found in \cite{epvm}). This closure is not tied up to the usual hydrodynamic gradient expansion, and it is nonlinear in a nonhydrodynamic variable $\gamma^{\mu\nu}$. The latter models the backreaction of $f$ (that could describe a highly nonequilibrium situation) on the hydrodynamic modes (that  relax much more slowly). In the EPP, $\gamma^{\mu\nu}$ is identified with the Lagrange multiplier of the variational problem whose solution gives the $f$ that extremizes the entropy production. Physically, one can think of the EPP as selecting the dynamics of $\gamma^{\mu\nu}$ (accounting for the backreaction of $f$ on $(\epsilon,u^\mu)$) in such a way as to extremize the production of entropy during the evolution of the system. This interpretation for $\gamma$ is rooted in nonequilibrium statistical physics, particularly in the study of fluctuations around stationary states \cite{prigo,jona,ruelle}.  

 The idea of a closure based on a variational principle whose Lagrange multipliers are macroscopic variables finds application in the formulation of Extended Thermodynamics \cite{extended}, which use as a variational principle the Maximum Entropy Principle (MEP) instead of the EPP.  On the other hand, the introduction of nonhydrodynamic variables into a macroscopic description derived from kinetic theory is not unique to the EPP. Related approaches are discussed in \cite{tanos1,tanos2,tanos3} for the MEP, in \cite{GKpaper}  where the moments of the collision term of the Boltzmann equation are treated as independent variables (see also \cite{karlin}), in \cite{mauricio,aniso,aniso2} where  AH is derived from an expansion of $f$ around an anisotropic state, in \cite{geroch} in the framework of divergence-type theories (see also \cite{dev,app,linking}), in \cite{denicol} where $f$ is expanded in moments to all orders in Knundsen number, and in \cite{visco} in the context of fluid models derived from the AdS/CFT correspondence.
 
We note that various effective theories derived from the MEP and the EPP have found applications in the study of electron transport through semiconductors \cite{tanos1,tanos2,tanos3,christen} and radiative heat transport in a photon gas \cite{christen2}. It was found that these effective theories are able to reproduce kinetic theory results better than standard hydrodynamic models.

\subsection{From kinetic theory to the NET}

We consider a relativistic kinetic theory of massless particles in flat space-time with signature $(-,+,+,+)$, and for simplicity neglect quantum statistics. 
The Boltzmann equations is 
\be 
p^\mu f(x,p)_{;\mu} = I_{{\rm col}}[f] 
\label{bolt}
\te
where $f$ is the one-particle distribution function, $I_{{\rm col}}[f]$ is the collision kernel, and a semicolon stands for a covariant derivative. We write the distribution function as 
\be 
f= f_0 [1+(1+f_0)\chi] \approx f_0 [1+\chi]
\label{fneq}
\te 
where $f_0 = {\rm exp}[-u_\nu p^\nu/T]$; the four velocity is obtained from the energy momentum tensor through the Landau-Lifshitz prescription $u_\mu T^{\mu\nu}  = \epsilon(T) u^\nu$ ($T$ is the temperature). We also define 
\be 
\lla \cdots \rla  = \int\:Dp\; f_0(\cdots)
\te  
where $Dp= d^4p \delta(p^2)/(2\pi)^3$. We shall base our developments on a relaxation time approximation for $I_{{\rm col}}$, suitably generalized to fulfill momentum conservation. We write the collision operator as 
\be 
I_{{\rm col}} = -\frac{1}{2\tau_r} R[FR \chi]
\te 
where $\tau_r$ is the relaxation time, $F=F(\omega)$ is an arbitrary function of energy $\omega=-p_\mu u^\mu$, and 
$R$ is an operator enforcing energy-momentum conservation $\lla p^\mu I_{{\rm col}} \rla=0$.  This form for $I_{{\rm col}}$ guarantees that the Second Law holds exactly. Moreover, it includes the well-known kinetic models of Marle (which is a relativistic generalization of the Bhatnagar-Gross-Krook model) and of Anderson-Witting \cite{cerc} as special cases if $F(\omega)=T$ and $F(\omega)=\omega$, respectively.

The stress-energy tensor is 
\be 
T^{\mu\nu}= T^{\mu\nu}_{0} + \Pi^{\mu\nu}
\te  
where 
\be 
T^{\mu\nu}_0 = \lla p^\mu p^\nu \rla
\te  
corresponds to a perfect fluid and 
\be 
\Pi^{\mu\nu}=\lla p^\mu p^\nu \chi \rla
\te 
encodes dissipative corrections. For a system of massless particles we have $T^{\mu\nu}_0 = \epsilon(u^\mu u^\nu +\Delta^{\mu\nu}/3)$, where $\Delta^{\mu\nu}=g^{\mu\nu}+u^\mu u^\nu$ is the spatial projector.
To write down the NET, we also introduce a nonequilibrium tensor $\gamma_{\mu\nu}$. The nonequilibrium tensor is symmetric and transverse, so in the rest frame has only spatial components $\gamma_{ij}$. Closure is obtained from the EPP:  

\be 
f=f_0 \bigg[1+\frac{\tau_r}{TF}p^i p^j \gamma_{ij}+
\frac{\tau_r^2}{2T^2F^2}p^i p^j p^l p^m \gamma_{ij}\gamma_{lm}\bigg]
\label{closfin}
\te 

Although the detailed derivation of Eq. (\ref{closfin}) can be found in \cite{linking,color}, for the sake of clarity it is convenient to briefly sketch the main logical steps that lead to this closure. The starting point is the variational equation whose solution gives the $f$ that extremizes the entropy production $S^\mu_{;\mu}[\chi]$ with the constraint (imposed by a Lagrange multiplier) that $\Pi^{\mu\nu}[\chi]$ take on known values. We note that  $S^\mu_{;\mu}$ and $\Pi^{\mu\nu}$ are functionals of $\chi$, the deviation of $f$ from equilibrium, and that the variation in the variational equation is performed with respect to $\chi$. The Lagrange multiplier enforcing the constraint on $\Pi^{\mu\nu}$ turns out to be proportional to $\gamma_{\mu\nu}$. Solving the variational equation (we stay at second order in the relaxation time) one finds a relation between $\chi$ and $\gamma_{\mu\nu}$, which, via Eq. (\ref{fneq}), leads directly to Eq. (\ref{closfin}). 

Once we have $f$ expressed in terms of our nonequilibrium tensor $\gamma^{ij}$, the rest of the calculation is standard: we can immediately compute the shear tensor and obtain the evolution equation for $\gamma^{ij}$ from the method of moments. We leave the latter task for the next subsection. 

The first order shear tensor is  $\Pi^{00}=\Pi^{k0}=0$, and
\be 
\Pi_1^{ij} =  2\eta  \gamma^{ij}
\label{pi1aux}
\te
The shear viscosity is
\be 
\eta = \frac{\tau_r T}{15}\left\langle \frac{\omega^4}{F} \right\rangle \equiv \frac{6\tau_\pi T}{15} \left\langle \frac{\omega^5}{F^2}\right\rangle^{-1}\left\langle \frac{\omega^4}{F} \right\rangle^2 
\label{eta}
\te 
and  $\tau_{\pi}$ is related by a constant to $\tau_r$
\be 
\frac{1}{\tau_\pi} \equiv \frac{6}{\tau_r}\left\langle \frac{\omega^5}{F^2}\right\rangle^{-1} \left\langle \frac{\omega^4}{F} \right\rangle 
\te  
The shear tensor at second order is 
\be
\Pi_{2}^{ij} = L [\gamma^{i}_k \gamma^{kj}-\frac{1}{3}\delta^{ij}\gamma^{lm}\gamma_{lm}]
\label{piphys}
\te 
with 
\be
L \equiv  \frac{\tau_r^2 }{45} \left\langle \omega^6/F^2 \right\rangle
\te 

It is interesting to remark that if we neglect the  second term in (\ref{closfin}), and set $\Pi_2^{\mu\nu}=0$, $F=T$ (corresponding to Marle's model) and $\tau_r=\eta/(\epsilon+P)$, where $P=\epsilon/3$ is the pressure, we recover Grad's quadratic ansatz given for a Boltzmann gas by \cite{cerc,revhydro2}
\begin{equation}
f=f_0\bigg(1+\frac{\Pi_{\mu\nu}p^\mu p^\nu}{2T^2(\epsilon+P)}\bigg)
\end{equation}
This shows that the closure discussed here is a nontrivial generalization of the commonly used quadratic ansatz, a fact that may have interesting consequences in the study of higher order anisotropy coefficients such as the hexadecapole flow $v_4$ (see e.g. \cite{luzum}).

\subsection{Evolution equations}

We now turn to the evolution equations of the NET. 

For the hydrodynamic variables $(\epsilon,u^\mu)$ these are simply $T^{\mu\nu}_{;\mu}=0$, i.e. energy-momentum conservation (exactly as in standard fluid dynamics).  The equation of motion for $\gamma_{ij}$ can be found from the transport equation (\ref{bolt}) by the method of moments. 
The second moment of the kinetic equation is 
\be 
\lla  p^\rho p^\sigma p^\mu f_{;\mu}[\gamma] \rla = \lla p^\rho p^\sigma I_{{\rm col}}[\gamma] \rla
\te 
where we write $[\gamma]$ to emphasize that $f_{;\mu}$ and $I_{{\rm col}}$ are functionals of $\gamma$ (recall that they are functionals of $\chi$, and the solution to the variational equation has provided us with a relation between $\chi$ and $\gamma^{\mu\nu}$). 
For simplicity, in what follows we will neglect the terms $\propto u_{\nu;\mu}\gamma^2$ and $\propto  \gamma_{lm} \gamma_{ij;\mu}$, which are expected to be very small. We get  
\be 
M^{\rho\sigma\mu\nu} u_{\nu;\mu} - \tau_r N_1^{ij\rho \sigma \mu} \gamma_{ij;\mu} 
= \frac{1}{2}\gamma_{ij}\bigg( N_1^{ij\rho\sigma}  + \frac{\tau_r}{T} N_2^{ijlm \rho \sigma} \gamma_{lm} \bigg)
\label{eomtemp}
\te 
where we have defined 
\be
M^{\rho\sigma\mu\nu} = N_0^{\rho\sigma\mu\nu} + \frac{\tau_r}{T}\gamma_{ij} N_1^{\rho\sigma\mu\nu i j}
\te 
and  
\be 
N_\alpha^{\rho\sigma\gamma \cdots \kappa} = \left\langle \frac{p^\rho p^\sigma p^\gamma \cdots p^\kappa}{F^\alpha}  \right\rangle
\label{Ns}
\te 
for brevity.

Eq. (\ref{eomtemp}) determines the evolution of the nonhydrodynamic variable $\gamma^{ij}$. We have already shown in \cite{dev} that if we expand $\gamma^{\mu\nu}$ to second order in velocity gradients, the above formalism goes over to SOFD \cite{hyd1,hyd2,hyd3}. We emphasize that the transport coefficients of the NET are chosen to be exactly the same as those of SOFD. This is because we want the NET to go over to SOFD when gradients are small. In doing so, we are neglecting higher order corrections to these transport coefficients (see \cite{denicol,denicolnew} for a discussion of this issue). Since ultimately we wish to apply the NET to describe the evolution of strongly coupled {\it dense} matter, for which kinetic theory does not hold, we believe that it is better to regard the transport coefficients of the NET as adjustable parameters. In any case, once the transport coefficients of SOFD are chosen, the NET has no free parameters. 

\section{Comparison to kinetic theory}
\label{comp}

We now compare the NET to SOFD and to kinetic theory. For the sake of simplicity, we focus on 0+1 boost invariant dynamics, which is a useful toy model of a heavy ion collision (see e.g. \cite{revhydro2}). All quantities are independent of space-time rapidity $\xi=\textrm{arctanh}(z/t)$ and transverse coordinates $(x,y)$ . Proper time is denoted by $\tau=\sqrt{t^2-z^2}$. 

It is convenient to introduce a second order transport coefficient 
\be 
\lambda_1 \equiv \frac{45 \tau_\pi \eta L}{\tau_r T^3}
\te 
which arises naturally in the context of SOFD \cite{revhydro2,hyd1,hyd2,hyd3}.  The equations of SOFD are (we write $\Pi\equiv \Pi^{\xi}_\xi$ and denote a time derivative with an overdot) \cite{revhydro2}
\begin{equation}
\begin{split}
\dot{\epsilon} &= -\frac{4}{3\tau}\epsilon + \frac{\Pi}{\tau} \\
\dot{\Pi} &= \frac{4\eta}{3\tau_\pi \tau} -\frac{\Pi}{\tau_\pi}-\frac{4}{3}\frac{\Pi}{\tau} 
-\frac{\lambda_1}{2\tau_\pi \eta^2}\Pi^2
\end{split}
\label{sofdeqs}
\end{equation}
In the NET, the evolution equation of $\gamma$ is (we write $\gamma \equiv \gamma^\xi_\xi$)
\begin{equation}
\dot{\gamma} = \frac{2}{3\tau_\pi \tau} -\frac{\gamma}{\tau_\pi}-\frac{4}{3}\frac{\gamma}{\tau} 
-\frac{\lambda_1}{\tau_\pi \eta}\gamma^2
\end{equation}
This equation must be supplemented with the relation between $\Pi$ and $\gamma$
\be 
\Pi = 2\eta \gamma + 2\lambda_1 \gamma^2
\te  
and with the equation for $\dot{\epsilon}$ given in (\ref{sofdeqs}). We emphasize that in the NET one first solves for the nonhydrodynamic variable $\gamma$ and then calculates $\Pi$ from it. A somewhat similar situation is encountered in the AH of \cite{mauricio,aniso,aniso2}, where the dynamical variable is (apart from the hydrodynamic ones) an anisotropy parameter from which $\Pi$ is computed {\it a posteriori}. 

To solve the Boltzmann equation we use the Lattice Boltzmann implementation recently developed in \cite{LB}. 
\begin{center}
\begin{figure}[htb]
\scalebox{0.47}{\includegraphics{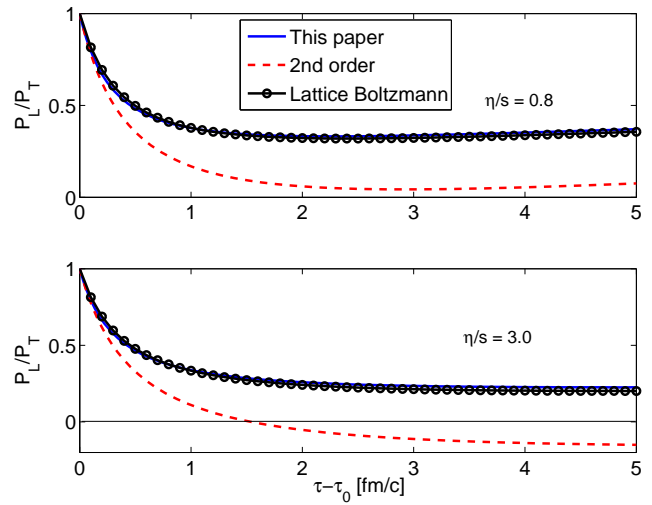}}
\caption{(Color online) Evolution of the pressure anisotropy $P_L/P_T=(P_0-\Pi)/(P_0+\Pi/2)$ for 0+1 boost invariant flow in the NET, SOFD and kinetic theory ($P_0$ is the equilibrium pressure), corresponding to the equilibrium initial condition. The transport coefficients are set to $\tau_\pi=6\eta/(sT)$ and $\lambda_1=\eta/(2\pi T)$. The upper panel corresponds to $\eta/s=0.8$, and the lower panel corresponds to $\eta/s=3.0$.}
\label{f1}
\end{figure}
 \end{center}

Figure \ref{f1} shows the evolution of the pressure anisotropy $P_L/P_T=(P_0-\Pi)/(P_0+\Pi/2)$ computed in the NET, SOFD and kinetic theory ($P_{0,L,T}$ are the equilibrium, longitudinal and transverse pressures), for the equilibrium initial condition $P_L/P_T=1$.
The results were obtained with the transport coefficients set to $\tau_\pi=6\eta/(sT)$, which is the value obtained from the kinetic theory of a Boltzmann gas \cite{revhydro2,hyd1,hyd2,hyd3}, and $\lambda_1=\eta/(2\pi T)$.  We show results for two different values of $\eta/s$, namely $\eta/s=0.8$ and $\eta/s=3.0$. We choose these values because they correspond to a typical value for the hadronic phase and a very large value used to test our model, respectively. The initialization time is $\tau_0 = 0.5$ fm/c, the initial temperature is $T_0 = 295$ MeV, and the initial energy density is $\epsilon_0 = 4.9$ GeV/fm$^3$. Our conclusions do not depend on the precise values of these initial values. 

\begin{center}
\begin{figure}[htb]
\scalebox{0.47}{\includegraphics{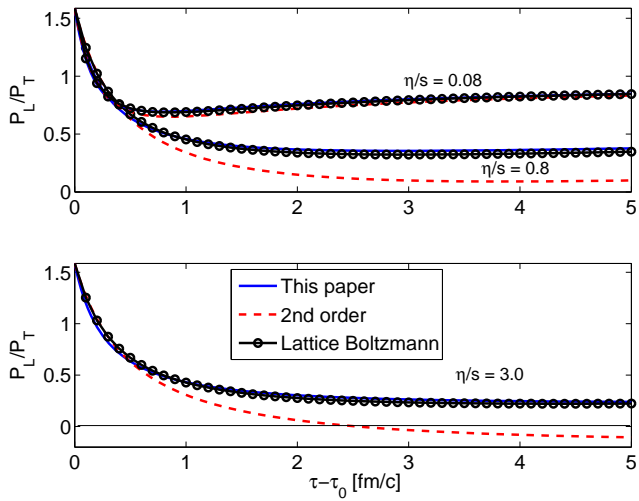}}
\caption{(Color online) Evolution of the pressure anisotropy $P_L/P_T=(P_0-\Pi)/(P_0+\Pi/2)$ for 0+1 boost invariant flow in the NET, SOFD and kinetic theory ($P_0$ is the equilibrium pressure), corresponding to a strong off-equilibrium initial condition. The transport coefficients are set to $\tau_\pi=6\eta/(sT)$ and $\lambda_1=\eta/(2\pi T)$. The upper panel corresponds to $\eta/s=0.08$ and $\eta/s=0.8$, and the lower panel corresponds to $\eta/s=3.0$.}
\label{f2}
\end{figure}
 \end{center}

From Figure \ref{f1} we see that the NET can track kinetic theory better than SOFD, even for a very large $\eta/s = 3.0$. On the other hand, the results obtained with SOFD do not agree with those of kinetic theory. For $\eta/s = 3.0$, the pressure anisotropy obtained from SOFD becomes negative, signaling the breakdown of this approach. It is worth mentioning that the agreement obtained between the NET and kinetic theory for such large values of $\eta/s$ is similar to that obtained from the developments of \cite{denicolnew} and \cite{third}. 
We note that for low values of $\eta/s \simeq 0.08$ (not shown in Figure \ref{f1}) the agreement between SOFD, the NET and kinetic theory is excelent (see Figure \ref{f2}). 

The equilibrium initial condition used to obtain Figure \ref{f1} does not correspond to the real situation encountered in a heavy ion collision. Matter created in such events exhibits very large momentum-space anisotropy at early times, so that a highly nonequilibrium initial condition for hydrodynamics (and the NET) seems more natural. Figure \ref{f2} shows the evolution of the pressure anisotropy for a strong off-equilibrium initial condition $P_L/P_T\simeq 1.5$. This value for $P_L/P_T$ corresponds to $|\Pi_0|\simeq 2 P_0/7$, which is close  to the constraint on the initial $\Pi_0$, namely $|\Pi_0|< P_0/3$, discussed in \cite{martedges} for the reliability of the hydrodynamic description of the evolution of the fireball. 

In Figure \ref{f2} we also show results for $\eta/s=0.08$, because it is not a priori evident that SOFD and the NET will be able to reproduce kinetic theory if a strong off-equilibrium initial condition is used. We find that indeed this is the case: for such low value of $\eta/s$, the agreement between SOFD, the NET and kinetic theory is excelent, even for very early times. In contrast, for $\eta/s = 0.8$ the SOFD deviates significantly from the result of kinetic theory, while the NET still tracks it very well during the entire evolution. For the very large value of $\eta/s=3.0$, the pressure anisotropy of SOFD becomes negative for $\tau-\tau_0 > 2$ fm/c. On the other hand, the NET is still able to track kinetic theory remarkably well at all times. Note, however, that slight discrepancies between the NET and kinetic theory start to show up.

As a check, we have used two other set of values for the second order transport coefficients, $\tau_\pi=3\eta/(sT)$ (strong coupling \cite{hyd1,hyd2,hyd3}) and $\lambda_1=\eta/(2\pi T)$, and $\tau_\pi=6\eta/(sT)$ and $\lambda_1=0$ (standard IS theory \cite{is1,is2}), and found that the NET also agrees better with kinetic theory than SOFD.

\section{Summary}
\label{summ}

In this paper we have tested the nonequilibrium model presented in \cite{linking,color} against second order fluid dynamics with regards to their ability to track the more fundamental kinetic theory description of a relativistic system. Our model assumes a closure that is nonlinear and does not rely on the gradient expansion, thus generalizing Grad's quadratic ansatz in a nontrivial way. The evolution equations for the macroscopic variables are obtained from the kinetic equation by the method of moments. 

By direct comparison to solutions of the Boltzmann equation for 0+1 boost invariant flow, we show that the effective theory describes the evolution of the pressure anisotropy better than standard fluid dynamics. The promising comparison to kinetic theory suggests that our model could be used to advantage  to study the early time evolution of matter created in heavy ion collisions, during which very large momentum-space anisotropies occur. Moreover, even for a strong off-equilibrium initial condition, our model can reproduce the results of kinetic theory for very large values of $\eta/s$. These  features make it an attractive choice to investigate the dynamics of the fireball produced in heavy ion collisions in numerical simulations including a realistic temperature-dependent $\eta/s$, which rises to large values in the hadronic phase before freeze out.

\begin{acknowledgements}
We thank Gast\~ao Krein for useful comments, and Paul Romatschke for making available the 0+1 Lattice Boltzmann code (Ref. \cite{LB}). This work has been supported in part by ANPCyT, CONICET and UBA under Project UBACYT X032 (Argentina).
 \end{acknowledgements}

\end{document}